\documentclass[aps,twocolumn,superscriptaddress,amssymb]{revtex4-1}

\usepackage{amsmath}

\usepackage{graphicx}
\usepackage{floatrow}
\usepackage[leftcaption]{sidecap}
\sidecaptionvpos{figure}{right}
\usepackage{amsmath,amsfonts,amsbsy,amssymb,array,accents,dsfont}
\usepackage{enumerate,array,latexsym,graphicx,mathrsfs,verbatim,psfrag}
\usepackage{bm} 
\usepackage[normalem]{ulem}
\usepackage{multirow}
\usepackage{cancel}
\usepackage{chngpage} 

\usepackage{booktabs} 
\usepackage[usenames]{color}
\usepackage[utf8]{inputenc}
\usepackage[english]{babel}
\usepackage{fancybox}
\usepackage[dvipsnames]{xcolor}

\usepackage{tikz}
\usepackage{pgfplots}
\usetikzlibrary{decorations.text,intersections, pgfplots.fillbetween}
\usetikzlibrary{math}
 \usetikzlibrary{snakes,3d,shapes.geometric,shadows.blur}
\usepackage{tikz-3dplot}

\usepackage{sidecap}

\sidecaptionvpos{figure}{c}

\usepackage{mciteplus}

\usepackage[colorlinks=true,      linkcolor=blue,      urlcolor=blue,      
            filecolor=blue,      citecolor=blue,       pdfstartview=FitH,     
						pdfpagemode=UseNone,      bookmarksopen=true]{hyperref}  
\usepackage[all]{hypcap}     

\makeatletter
\renewcommand{\p@subsection}{}
\renewcommand{\p@subsubsection}{}
\makeatother

\newcommand{\captionfonts}{\small}
\makeatletter  
\long\def\@makecaption#1#2{%
  \vskip\abovecaptionskip
  \sbox\@tempboxa{{\captionfonts #1: #2}}%
 \ifdim \wd\@tempboxa >\hsize
    {\captionfonts #1: #2\par}
  \else
    \hbox to\hsize{\hfil\box\@tempboxa\hfil}%
  \fi
  \vskip\belowcaptionskip}
\makeatother   

\DeclareMathSymbol{\medhatsym}{\mathord}{largesymbols}{"62} 

\DeclareMathSymbol{\medtildesym}{\mathord}{largesymbols}{"65}

\newcommand{\comm}[1]{} 

\def\IR{\mathbb{R}}

\def\({\left(}
\def\){\right)}
\def\[{\left[}
\def\]{\right]}

\def\One{{\hbox{ 1\kern-.8mm l}}}

\def\barray{\begin{array}}
\def\earray{\end{array}}
\def\be{\begin{equation}}
\def\ee{\end{equation}}
\def\bea{\begin{eqnarray}}
\def\eea{\end{eqnarray}}
\def\bal{\begin{align}}
\def\eal{\end{align}}

\def\-{\,-\,}
\def\={\,=\,}
\def\+{\,+\,}

\numberwithin{equation}{section} 

\makeatletter
\g@addto@macro\bfseries{\boldmath}
\makeatother

\definecolor{cardinal}{rgb}{0.6,0,0}
\definecolor{darkgreen}{rgb}{0,0.4,0}
\definecolor{purple}{rgb}{0.5, 0, 0.5}
\definecolor{golden}{rgb}{0.92, 0.7, 0}
\definecolor{midnight}{rgb}{0, 0, 0.5}
\definecolor{darkblue}{rgb}{0, 0, 0.8}

\def\IR{\mathds{R}}

\usetikzlibrary{positioning,calc,fadings,decorations.pathreplacing}

\tikzset{
 diffuse color/.initial = black,                       
}

\tikzset{
 linear opacity/.initial=0.5,                          
 linear stroke/.style = {                              
   preaction={                                         
     draw=\pgfkeysvalueof{/tikz/diffuse color},        
     line width = (2.0-#1)*\pgflinewidth,              
     opacity=\pgfkeysvalueof{/tikz/linear opacity},white}},  
 diffuse gradient/.style={                             
   draw = none,                                        
   linear opacity=#1,                                  
   linear stroke/.list={0.0,#1,...,1.0}},              
 diffuse gradient/.default=1,                          
}

\tikzset{
 non-linear stroke/.style = {                          
   preaction={                                         
     draw=\pgfkeysvalueof{/tikz/diffuse color},        
     line width = (2.0-#1)*\pgflinewidth,              
     opacity=#1,white}},                                     
 diffuse falloff/.style={                              
   draw = none,                                        
   non-linear stroke/.list={0.0,#1,...,1.0}},          
 diffuse falloff/.default=1,                           
}
 
\tikzset{%
  >=latex, 
  inner sep=0pt,%
  outer sep=2pt,%
  mark coordinate/.style={inner sep=0pt,outer sep=0pt,minimum size=3pt,
    fill=black,circle}%
}

\usepackage{titlesec}
\titlespacing*{\section}
{0pt}{0.5cm}{0.2cm}
\titlespacing*{\subsection}
{0pt}{0.25cm}{0.1cm}

\begin{document}

\title{Topological Stars and Black Holes }

\author{Ibrahima Bah}
\email{iboubah@jhu.edu}
\author{Pierre Heidmann}
\email{pheidma1@jhu.edu}
\affiliation{Department of Physics and Astronomy, Johns Hopkins University, 3400 North Charles Street, Baltimore, MD 21218, USA}

\begin{abstract}

We study smooth bubble spacetimes in five-dimensional Einstein-Maxwell theory that resemble four-dimensional magnetic black holes upon Kaluza-Klein reduction. We denote them as Topological Stars since they have topological cycles supported by magnetic flux. They can be macroscopically large compared to the size of the Kaluza-Klein circle and could describe qualitative properties of microstate geometries for astrophysical black holes.  We also describe five-dimensional black strings without curvature singularity, the interior caps as a two-dimensional Milne space with a bubble.

\end{abstract}


\maketitle

Black holes have provided the basic theoretical laboratory for exploring quantum theories of gravity. However, the excitement for black holes are no longer theoretical. Collisions of black holes can be observed by the LIGO collaboration \cite{Abbott:2016blz}, and their near environment can be imaged \cite{Akiyama:2019cqa}. In this new age of astronomy, it is interesting to wonder whether theoretical results can lead to new observables. The main theoretical questions are about the nature of the degrees of freedom that can resolve black-hole singularities, and how they account for microstates of the Bekenstein-Hawking entropy. 

A priori, microstates are quantumly.  However, some can be sufficiently coherent to admit a classical description as ultra-compact objects. In four dimensions, such microstates can be contrasted with compact objects like gravastars \cite{Mazur:2001fv} or boson stars \cite{Schunck:2003kk}.  These constructions require exotic matter beyond General Relativity (GR) and strong fine-tunings for which the UV origin is unclear.

In String Theory, a large number of horizonless supersymmetric ``microstate geometries'' has been constructed. These resemble the black hole up to Planck-length distance above its horizon. They are classical fuzzballs \cite{Mathur:2005zp} and require fluxes on non-trivial topological cycles, called bubbles, at the vicinity of the would-be horizon. This is the only viable mechanism supporting vast amounts of horizon-scale structure \cite{Gibbons:2013tqa}. However, most of the solutions, from the first constructed \cite{Bena:2004de} to the large classes \cite{Bena:2007kg,Bena:2016ypk,Heidmann:2019xrd}, except for a few \cite{Bena:2009qv}, correspond to unrealistic black holes for astrophysics. All constructions require sophisticated machinery in supergravity. 

In this letter, we aim to fill the gap between beyond-GR toy models that are convenient for phenomenology and top-down string-theory microstate geometries. We study simple smooth bubble geometries in four dimensions times a circle.  They look like non-extremal non-supersymmetric static black holes in four dimensions upon Kaluza-Klein reduction. Their five-dimensional nature is manifest in the region near the horizon.  We refer to the solutions as ``Topological Stars" \cite{Bena:2013dka}.  

The solutions are constructed in the simplest possible framework: five-dimensional Einstein-Maxwell theory.  By Kaluza-Klein reduction, they are solutions of an Einstein-Maxwell-Dilaton theory in four dimensions. 

The key ingredient is to use topological cycles, as in microstate geometries, to replace the horizon by some degeneracy of the extra circle.  A simple way to construct them is to consider a double Wick rotation of a black string where the time direction maps to a circle direction and vice versa.  The original horizon then becomes a region where a circle shrinks, thereby corresponding to a bubble of nothing \cite{Witten:1981gj}. The spacetime in these cases are massless and unstable in five dimensions.  Massive bubbles can be obtained by adding suitable magnetic flux, which necessitates Maxwell fields, as discussed in \cite{STOTYN2011269}.

The radius of the extra dimension, $R_y$, must be small in the asymptotic region in order to have effective four-dimensional physics.  It must also be larger than the string scale to avoid quantum instabilities \cite{Adams_2005}.  Regularity of magnetic bubbles fix their size to be smaller than $R_y$ \cite{STOTYN2011269}.  In this letter, we study more interesting and nontrivial regularity conditions than in \cite{STOTYN2011269} by allowing orbifold fixed points, and their classical resolution as Gibbons-Hawking bubbles.  This allows for the size of the topological star to take any value independent from $R_y$. The resolution hints at additional degrees of freedom for a richer class of microstate geometries.  

We also discuss the properties of the magnetic black strings in the five-dimensional Einstein-Maxwell theory with respect to which our smooth solutions should be compared to. They are interesting on their own right.  In addition to a horizon, the interior caps as a two-dimensional Milne space with a bubble thereby resolving the would be curvature singularity.  

Our construction allows for a more qualitative understanding of bubbles as microstate geometries. The solutions can be used for the study of black hole astrophysics and gravitational-wave physics. Moreover, we can embed them in string theory and explore their microstate nature from a top-down perspective.

In section \ref{sec:4+1dsol}, we detail the framework and our generating technique to obtain a class of spherically-symmetric solutions containing both topological stars and black strings. We describe them in detail in section \ref{sec:Bubblephase} and \ref{sec:BHphase} respectively. We discuss the phase space in section \ref{sec:phasespace} and some generalizations in section \ref{sec:generalizationDdim}.

\section{The Solutions in $4+1$ dimensions}
\label{sec:4+1dsol}

In this section, we review the class of solutions of \cite{Miyamoto:2006nd,STOTYN2011269}. A bubbling topology is induced by a shrinking circle while other compact directions keep a finite size. A mass can be generated by turning on flux  \`a la microstate geometries.  We consider solutions of five dimensional Einstein-Maxwell theory,
\begin{equation}
S = \int d^5x\sqrt{-g} \left(\frac{1}{2 \kappa_5^2} R - \frac{1}{4} F_{\mu\nu} F^{\mu \nu} \right), 
\label{eq:EinsMaxAc5}
\end{equation} 
that asymptote to four-dimensional Minkowski times a circle and where $\kappa_5 $ is the gravitational coupling, $F$ is the field strength. A spherically-symmetric metric ansatz with a magnetic flux leads to
\begin{equation}
\begin{split}
ds^2 &\= -f_\text{S}(r) \,dt^2 + f_\text{B}(r) \,dy^2 + \frac{dr^2}{h(r)} + r^2 \,d\Omega_2^2,\\
F &\= P \, \sin\theta\,d\theta \wedge d\phi \,,
\end{split} \label{eq:metAns}
\end{equation} where $d\Omega_2^2 = d\theta^2+\sin^2\theta\, d\phi^2$.  The coordinate $y$ parametrizes a circle with period $2\pi R_y$. First, we consider the vacuum solutions, $P=0$:
\begin{align}
1.& \quad f_\text{S}(r) = h(r) = 1 - \frac {r_\text{S}}{r}, \quad f_\text{B}(r) =1 \,,\label{BH} \\
2.&  \quad f_\text{B}(r) = h(r) = 1 - \frac {r_\text{B}}{r}, \quad f_\text{S}(r)=1\,.  \label{BL}
\end{align}
The first one, \eqref{BH}, is a product of a four-dimensional Schwarzschild black hole with a circle. It has a horizon at $r=r_\text{S}$ where the timelike Killing vector, $\partial_t$, shrinks. The second solution, \eqref{BL}, is a smooth massless solution that corresponds to static bubble of nothing at $r=r_\text{B}$ where the spacelike Killing vector, $\partial_y$, shrinks.  The two solutions are related by double Wick rotation $(t,y,r_\text{S},r_\text{B}) \rightarrow (i y, i t,r_\text{B},r_\text{S})$. 

We superpose both vacuum solutions by imposing the double Wick rotation as a symmetry \cite{Miyamoto:2006nd,STOTYN2011269}
\begin{align} \
f_\text{B}(r) &= 1-\frac{r_\text{B}}{r}, \qquad f_\text{S}(r)= 1-\frac{r_\text{S}}r, 
\end{align}
and the equations are solved providing
\begin{equation}
h(r) = f_\text{B}(r) \,f_\text{S}(r), \qquad P =\pm \frac{1}{\kappa_5} \sqrt{\frac{3\, r_\text{S} r_\text{B}}{2}}\,.
\end{equation}


There are two coordinate singularities, a horizon at $r=r_\text{S}$ and a degeneracy of the $y$-circle at $r=r_\text{B}$. Depending on the order, they either correspond to a massive magnetic bubble for $r_\text{B} > r_\text{S}$ or a magnetic black string for $r_\text{S} \geq r_\text{B}$.

By KK reduction along $y$, they are solutions of an Einstein-Maxwell-dilaton system  \footnote[1]{The general KK reduction of \eqref{eq:EinsMaxAc5}, for generic solutions 
$$ 
ds_5^2 = e^{-4 \Phi} (dy+\mathcal{A})^2 + e^{2\Phi} ds_4^2\,, \quad F = F_{(1)} + d\chi \wedge (dy+\mathcal{A}),
$$
gives the action 
\begin{equation*}
\begin{split}
S_4=\int \mathrm{d}^{4} x \sqrt{-g} &\left(\frac{1}{2 \kappa_4^2} R_{4} - \frac{3}{\kappa_4^2}\, \partial_a \Phi\, \partial^a \Phi - \frac{e^{4\Phi}}{2e^2} \partial_\mu \chi \partial^\mu \chi  \right. \\
&\left. -\frac{e^{-6\Phi}}{8 \kappa_4^2} \mathcal{F}_{\mu \nu} \mathcal{F}^{\mu \nu} -\frac{e^{-2\Phi}}{2e^2} {F_{(1)}}_{\mu \nu} F_{(1)}^{\mu \nu} \right),
\end{split}
\end{equation*} where $\mathcal{F}=d\mathcal{A}$. In \eqref{eq:Action4d}, we did not consider all KK degrees of freedom that are trivial for our solutions.}
\begin{equation} \small
S_4=\int \mathrm{d}^{4} x \sqrt{-g} \left(\frac{1}{2 \kappa_4^2} R_{4} - \frac{3}{\kappa_4^2}\, \partial_a \Phi\, \partial^a \Phi -\frac{e^{-2\Phi}}{4e^2} F_{\mu \nu} F^{\mu \nu} \right),
\label{eq:Action4d}
\end{equation} 
where $\kappa_4^2 = \frac{\kappa_5^2}{2\pi R_y}$ and $e^2 = \frac{1}{2\pi R_y}$ given by
\begin{align}
e^{2\Phi} &= f_\text{B}^{-1/2}, \qquad F = \pm \frac{e}{\kappa_4} \sqrt{ \frac{3r_\text{B} r_\text{S}}{2}} \sin \theta d\theta \wedge d\phi \nonumber \\
ds^2_\text{KK} &= f_\text{B}^{1/2} \left[ - f_\text{S}\, dt^2 + \frac{dr^2}{f_\text{B} f_\text{S}} + r^2\, d\Omega^2 \right].  
\end{align}
The four-dimensional ADM mass,  $M$, and the magnetic charge, $Q_m$, are
\begin{equation}
 M= 2\pi \left(\frac{2\,r_\text{S}+r_\text{B}}{\kappa_4^2} \right) \,,\qquad Q_m^2 =  \frac{3}{2} \,\frac{r_\text{B} r_\text{S}}{\kappa_4^2} .
\label{eq:ADMmass&Qm} 
\end{equation}

\section{The Topological Star}
\label{sec:Bubblephase}

\begin{figure}
\floatbox[{\capbeside\thisfloatsetup{capbesideposition={right,center},capbesidewidth=3cm}}]{figure}[\FBwidth]
{\caption{Penrose diagram of the topological star. The spacetime
ends as a smooth bubble at $r = r_\text{B}$.}\label{fig:PenroseBubble}}
{\includegraphics[height=3.2cm]{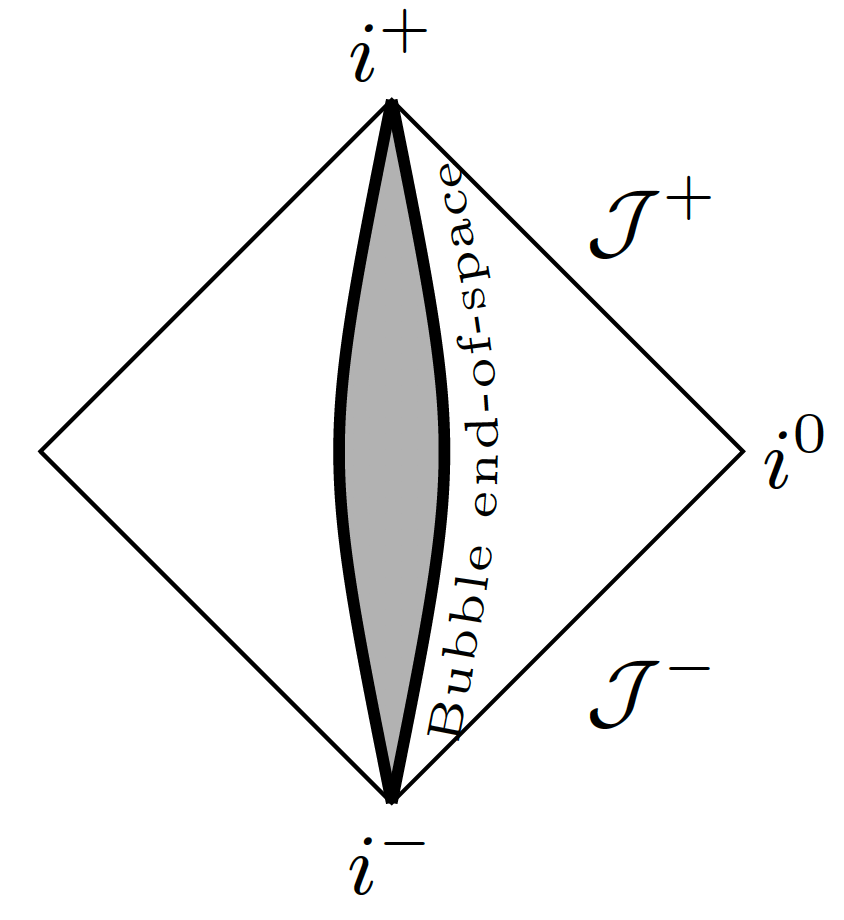}
}
\end{figure}

For $r_\text{B} > r_\text{S}$, the $y$-circle shrinks at $r=r_\text{B}$ providing an end to spacetime. We describe the local metric taking $\rho^2 =\frac{ 4(r-r_\text{B})}{r_\text{B}-r_\text{S}}$ with $\rho \to 0$,
\begin{align}
ds^2_5  &= - \frac{r_\text{B}- r_\text{S}}{r_\text{B}}\,dt^2 + r_\text{B}^2\, ds^2_4 \,,\label{eq:bubblemetricBP} \\
ds^2_4 &= d\rho^2 + \frac{r_\text{B}-r_\text{S}}{4 \,r_\text{B}^3}\,\rho^2 \,dy^2+d\Omega_2^2  \,.\nonumber
\end{align}  The line element, $d\Omega_2^2$, describes a two-sphere of radius $r_\text{B}$ while the $(\rho,y)$ subspace makes a flat surface. In general, we can allow for a conical defect and take the constant time-slices as $\mathbb{R}^2/\mathbb{Z}_k \times S^2$, $k\in \mathbb{Z}_+$. The Penrose diagram is summarized in Figure \ref{fig:PenroseBubble}.  
The orbifold condition relates the radius of the $y$-circle with the parameters as
\begin{equation}
k^2 R_y^2 = \frac{4\, r_\text{B}^3}{r_\text{B} -r_\text{S}} \quad \leftrightarrow \quad r_\text{S} = r_\text{B} \left(1-4 \frac{r_\text{B}^2}{k^2 R_y^2} \right).
\label{eq:quantization}
\end{equation} In general, $R_y$ is considered as a fixed asymptotic quantity, and therefore we interpret the regularity condition as fixing $r_\text{S}$.  We write the mass and charge as
\begin{equation} 
M = \frac{2\pi r_\text{B}}{\kappa_4^2}  \left(3-8 \frac{r_\text{B}^2}{k^2R_y^2} \right), \quad  Q_m^2 = \frac{3 r_\text{B}^2  }{2\kappa_4^2 } \left(1-4 \frac{r_\text{B}^2}{k^2R_y^2} \right).  \label{eq:bubbleMQ} \nonumber
\end{equation}  The radius of the bubble has an upper bound $2r_\text{B} \leq k R_y$.  At equality, the charge vanishes and the solution is simply a vacuum bubble of nothing.  

For macroscopic objects, we must have a bubble with large $k$. For $k=1$, the spacetime is smooth. For $k\geq 2$, there is a source corresponding to a conical defect localized at $r=r_\text{B}$, wrapping the bubble.

One can ask if a geometric transition can replace the source with topology.  We consider a more suitable choice of coordinates, $(\phi = \chi -\frac{1}{k}\beta, \; y =  R_y \beta)$ to  study the orbifold of the near-bubble metric \eqref{eq:bubblemetricBP} \cite{Bah:2019jts},
\begin{align}
ds^2_4 &=  d\rho^2 + d\theta^2 + \frac{\sin^2 \theta\, \rho^2}{R^2_\beta}  d\chi^2 + \frac{1}{k^2} R_\beta^2 D\beta^2 \nonumber \\
D\beta &= d\beta - k \frac{\sin^2 \theta}{ R_\beta^2} d\chi, \quad R_\beta^2 = \rho^2 +\sin^2 \theta. 
\end{align} The space is an $S^1_\beta$-bundle over a three-dimensional base given by $(\rho, \theta, \chi)$.  At $\rho=0$, the $\chi$-circle shrinks and $(\theta, D\beta)$ forms a $S^2/\mathbb{Z}_k$ bubble.  

The connection of the $S^1_\beta$-bundle has monopole charges $k$ at the north pole, $(\rho =0, \theta=0)$, and $-k$ at the south pole, $(\rho =0, \theta=\pi)$. Each pole corresponds to a single-center Gibbons-Hawking (GH) space, $\mathbb{R}^4/\mathbb{Z}_k$, with charge $k$.  The orbifold singularity can be classically resolved by splitting the GH center into $k$ centers of charge one where the $\beta$-fiber shrinks. There are two-cycles between any two centers and the singularity is replaced by a smooth space with $k-1$ bubbles. They exist in a limit where their characteristic size  is much smaller than the original bubble.

\begin{figure}[ht]
\begin{center}
\includegraphics[height=4.3cm]{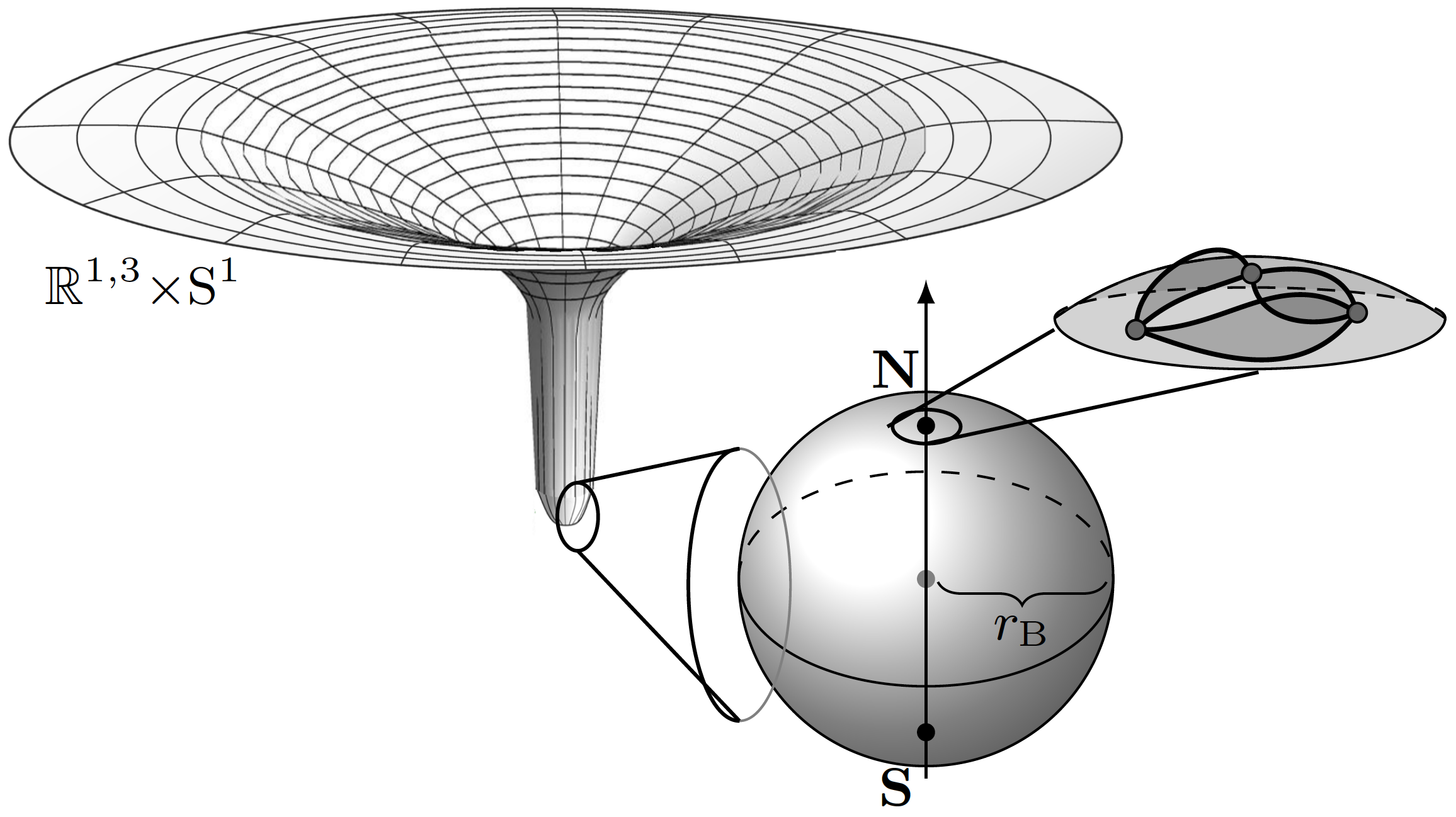}
\caption{Schematic description of a generic topological star.}
\label{fig:Schematic}
\end{center}
\end{figure}

Our one-bubble solution is a highly symmetric spacetime that highlights a much richer space of states via the classical resolution of its orbifold singularity (see Fig.\ref{fig:Schematic}).  We can turn on magnetic fluxes along each small bubble. We expect large degrees of freedom on the fluxes while fixing the conserved charge which implies a larger class of multi-bubble solutions without spherical symmetry.

\section{The Black String}
\label{sec:BHphase}
When $r_\text{S} > r_\text{B}$, the first coordinate singularity is the horizon at $r=r_\text{S}$. It has a S$^1\times$S$^2$ topology where the S$^2$ has a radius $r_\text{S}$ while the S$^1$ has a radius $\frac{\sqrt{r_\text{S}-r_\text{B}}}{\sqrt{r_\text{S}}} R_y$. 

\begin{figure}[ht]
\begin{center}
\includegraphics[height=3.7cm]{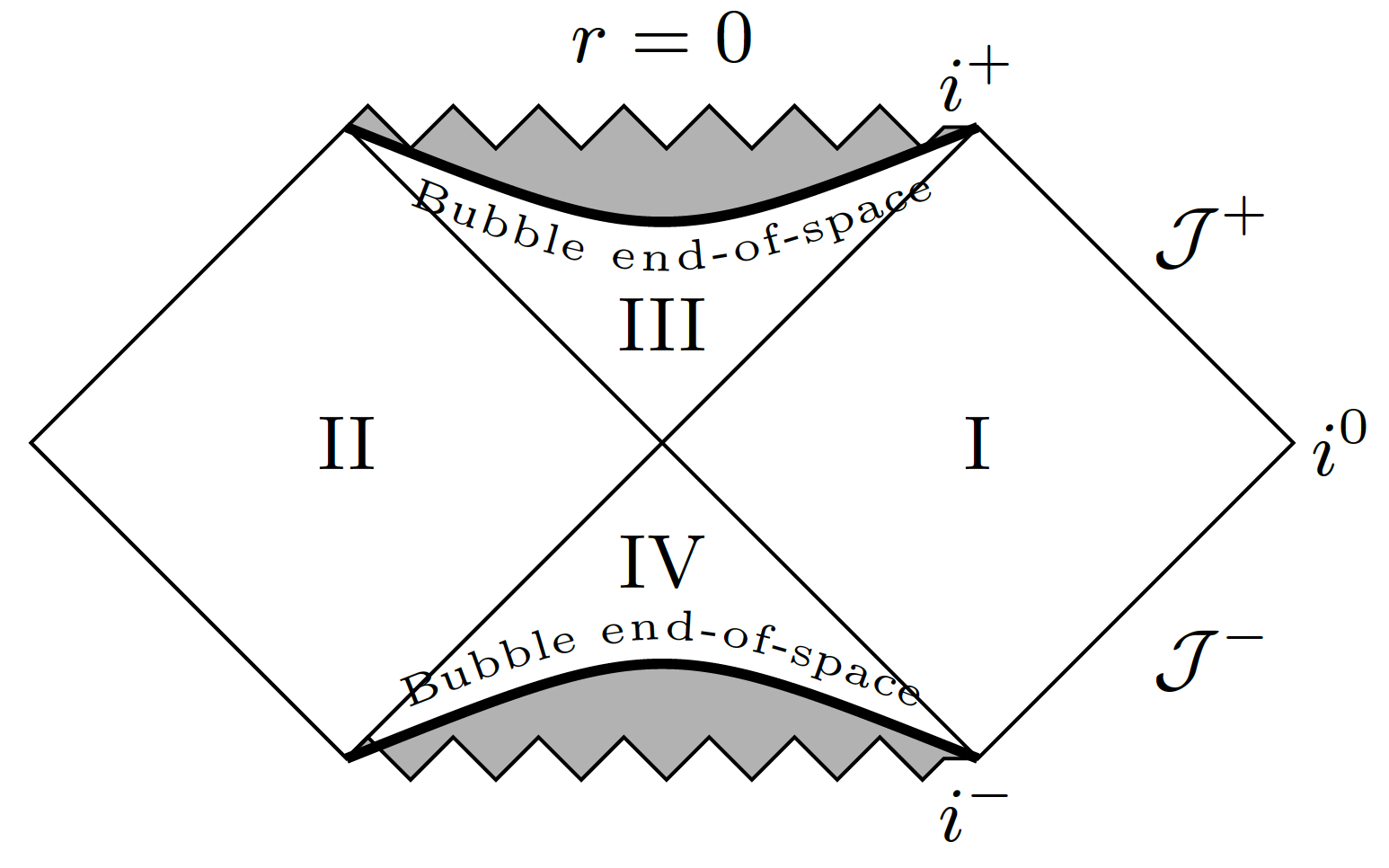}
\caption{Penrose diagram of the black string.}
\label{fig:PenroseBH}
\end{center}
\end{figure}

The coordinate singularity at $r=r_\text{B}$ is hidden behind the horizon. In this region, the spacelike Killing vector $\partial_y$ shrinks, thereby defining a bubble behind the horizon.  The causal structure of the spacetime is depicted by the Penrose diagram Fig.\ref{fig:PenroseBH}.

Unlike \eqref{eq:bubblemetricBP}, this bubble is a timelike surface and sits at the origin of a two-dimensional Milne space \cite{HOROWITZ199191} described by the $(\rho,y)$ subspace.  It is defined as the quotient of $\IR^{1,1}$ by a boost $\gamma^2  = \frac{r_\text{S}-r_\text{B}}{4 r_\text{B}^3} R_y^2$, and corresponds to cones in $\IR^{1,1}$ connected at their tips. The spacetime has no curvature singularity or closed timelike curves. However, geodesics with $y$-momentum are singular and experience a Cauchy horizon. Geodesics without $y$-momentum can be uniquely extended past the tip of the cone.  This suggests that if we restrict to energies and particles bellow the KK scale, we can connect two of the black hole geometries in Fig \ref{fig:PenroseBH} by identifying their bubble regions.  This describes a new class of possibly stable wormholes bellow the KK scale, and deserves further study. 

From an external four-dimensional perspective, we have a magnetic black hole of mass and charge given by \eqref{eq:ADMmass&Qm}, its horizon is at $r=r_\text{S}$ and its entropy is $\kappa_4^2\,S =8\pi^2 \sqrt{r_\text{S}^3\, (r_\text{S}-r_\text{B})}\,$.

In the limit $r_\text{B} \to r_\text{S}$, the bubble approaches the horizon and the solution is an extremal black string when $r_\text{B}=r_\text{S}=m$. Considering $r=\rho^2+m$, the metric is
\begin{equation} 
\begin{split}
ds^2 = & \rho^2 \left(\rho^2 +m \right)^{-1}\,\left[ -dt^2 +dy^2\right] \\
&+ \rho^{-2} \left(\rho^2+m \right)^2\,\left[4 d\rho^2 + \rho^2 \, d\Omega_2^2 \right]\,.
\end{split} \label{eq:extremalBH}
\end{equation}
The near-horizon region, $\rho \to 0$, corresponds to an AdS$_3\times$S$^2$. The radius of the AdS$_3$ and the S$^2$ are $2m$ and $m$ respectively.

\section{Phase Space}
\label{sec:phasespace}

\begin{figure}[ht]
\begin{center}
\includegraphics[height=4.2cm]{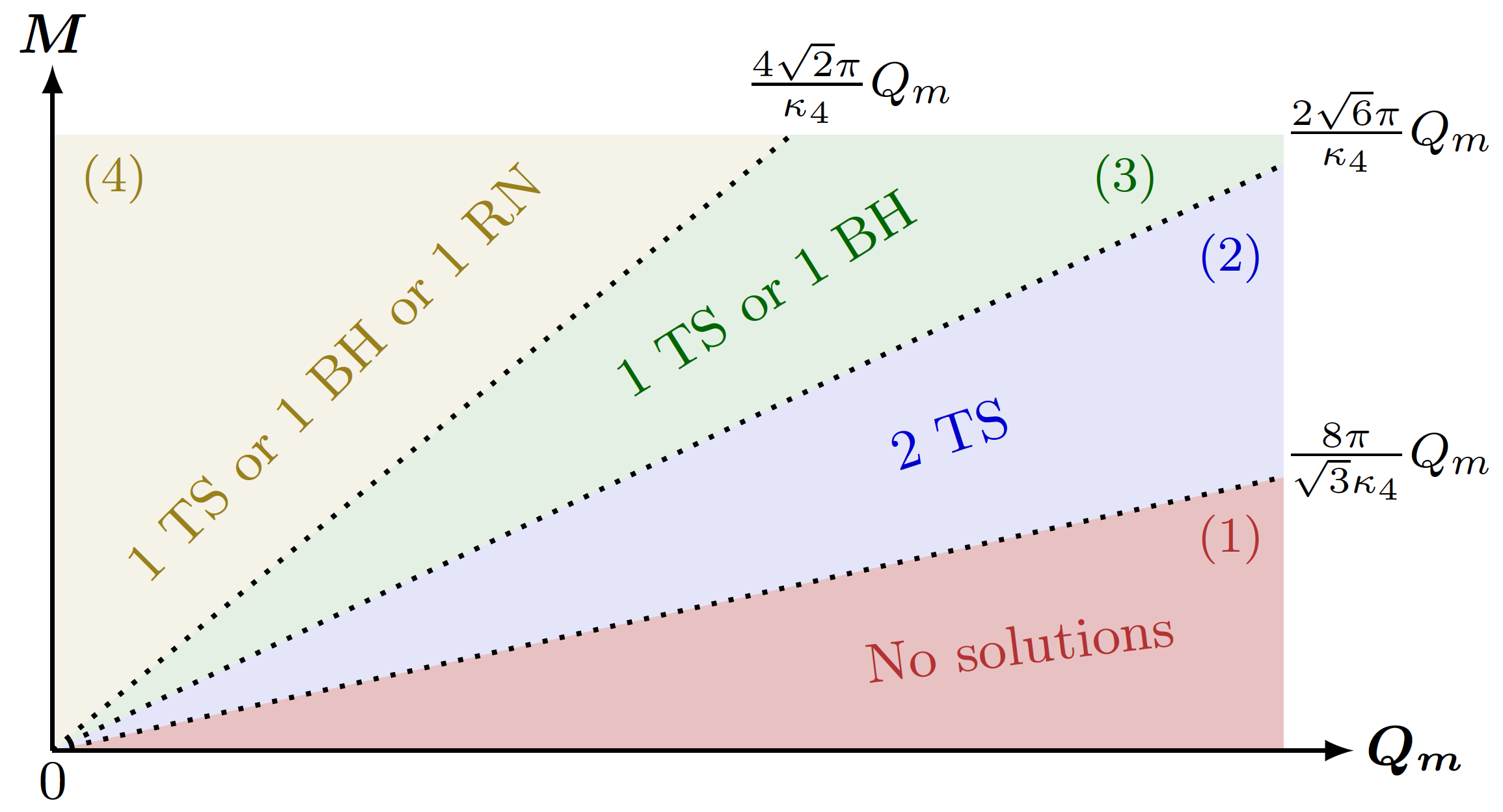}
\caption{Phase space of spherically-symmetric solutions. ``TS'', ``BH'' or ``RN'' stand for the topological star of section \ref{sec:Bubblephase}, the black string of section \ref{sec:BHphase} and the magnetic Reissner-Nordstr\"om respectively.}
\label{fig:phasespace}
\end{center}
\end{figure}

We consider the phase space of spherically-symmetric solutions with fixed ADM mass $M$, charge $Q_m$ and radius $R_y$.  The solutions are given in terms of two parameters $(r_\text{B}, r_\text{S})$ related to $(M, Q_m)$ as in \eqref{eq:ADMmass&Qm}.  In general, there are two solutions for $(r_\text{B},r_\text{S})$ given $(M, Q_m)$.  In Fig.\ref{fig:phasespace}, we plot the ranges for when topological stars and black strings exist.  On the same plot, we include the allowed range for magnetic Reissner-Nordstr\"om black holes (RN) in four dimensions. Note that it is not a solution of \eqref{eq:Action4d} and must be seen as an illustrative comparison. Its horizon radius is $8\pi R_{RN} = \kappa_4^2 \left( M + \sqrt{\kappa_4^2 M^2 - 32 \pi^2 Q_m^2} \right)$.

For small mass (region (2) in Fig.\ref{fig:phasespace}), both solutions of $(r_\text{B},r_\text{S})$ correspond to topological stars.  However, they exist for different choices of $R_y$ following from the regularity condition in \eqref{eq:quantization}. As the mass is increased at fixed $Q_m$,  the bubble with the smaller radius disappears as $r_\text{B} \to r_\text{S}$, and we hit the extremality line, $\kappa_4 M = 2\pi \sqrt{6} Q_m$, which corresponds to the extremal black string solution in \eqref{eq:extremalBH}.  In region (3), we have the non-extremal black string of the previous section and a topological star with the same mass and charge.

As the mass is further increased, we hit the cosmic censorship bound $\kappa_4 M = 4\pi  \sqrt{2} Q_m$ for magnetic RN black holes.  Beyond this line, topological stars, black strings and magnetic RN exist in the same regime. It is worth to compare the size of each object for given $(M,Q_m)$ in this regime. We denote $R_{BH}$ and $R_{TS}$ the radius of the black hole solution and the topological star respectively. We have
\begin{equation}
R_{BH} \sim R_{RN}\quad \text{and} \quad R_{TS} = 2 \,R_{BH}\,.
\end{equation}

The take-away message in the context of black hole microstates is that, even if we restrict to spherically-symmetric solutions, we have smooth bubble solutions that have the same charge and mass as the non-extremal four-dimensional black holes. It is appropriate to interpret this solution as a microstate of the thermal ensemble given by the Bekenstein-Hawking entropy. It is already surprising that such a state can be built with spherical symmetry. Moreover, because the topological star is twice as large as the size of its corresponding black hole, it is a rather atypical state. This is a common story for microstate geometries. Having solutions that scale very close to the horizon requires to consider multi-bubble solutions, that is, to break the spherical symmetry.

\section{Generalization}
\label{sec:generalizationDdim}
We explore interesting generalizations in a companion paper \cite{bigpaper}. We generalize to $D+1$ dimensions by imposing a similar double Wick rotation symmetry on the $D$-dimensional Schwarzschild-Tangherlini solution \cite{Tangherlini:1963bw} with an extra dimension. We also enrich the theory with higher-form fields and introduce an electric line charge.

The generalized solutions in Einstein-Maxwell theory in $D+1$ dimensions with a two-form electric and $(D-3)$-form magnetic gauge fields are
\begin{align}
ds^2_{D+1} &= - f_\text{S}(r) dt^2 + f_\text{B}(r) dy^2 + \frac{dr^2}{f_\text{S}(r) f_\text{B}(r)} + r^2 d\Omega^2_{D-2} \,, \nonumber \\
f_\text{S}(r) &= 1-\left(\frac{r_\text{S}}{r}\right)^{D-3} , \quad f_\text{B}(r) =1-\left(\frac{r_\text{B}}{r}\right)^{D-3}. \nonumber
\end{align}
They are supported by the fields 
\begin{equation}
F^{(m)} = P \, dV_{S^{D-2}} \,, \quad F^{(e)} = \frac{Q}{r^{D-2}} \, dt\wedge dr\wedge dy \,,
\end{equation}
where the magnetic and electric charges satisfy
\begin{equation}
P^2+Q^2 = \frac{(D-3)(D-1) \, r_\text{S}^{D-3} r_\text{B}^{D-3}}{2\,\kappa_{D+1}^2}\,.
\end{equation}
Here $d\Omega_{D-2}^2$ and $dV_{S^{D-2}}$ are the line element and the volume form of a unit $(D-2)$-sphere.  This class of solutions has a similar phase space as in four dimensions. 

We can embed the solutions in type IIB string theory, for $D = 4,5$, by considering a T$^{9-D}$ compactification with the NS-NS fields turned off and the fluxes $(F^e,F^m)$ arising from the R-R fields.  Having a rigid T$^{9-D}$ requires to fix the electric charge with respect to the magnetic charge. The $D=5$ case is a special subject in the microstate geometry program to construct smooth solutions that look like the extremal or non-extremal three-charge black hole \cite{Bena:2004de,Bena:2007kg,Bena:2015bea,Heidmann:2019xrd,Jejjala:2005yu,Bossard:2014ola}. Our methods allow tthe construction of the first bubbling geometries in the same regime as the static non-extremal D1-D5 black hole.

Moreover, we can obtain a larger class of bubbling solutions by breaking spherical symmetry. As observed in section \ref{sec:Bubblephase}, more multi-bubble solutions exist when the $y$-circle is twisted over the $\phi$-circle. This classically resolves the conical defects at the poles of the bubble. As a first step, we can construct axially-symmetric multi-bubble solutions with flux that generalize the Weyl formalism of \cite{Weyl:book}, which has allowed for the study of vacuum ``multi-rod'' solutions in 4+1 dimensions \cite{Emparan:2001wk,Elvang:2002br,Emparan:2008eg,Charmousis:2003wm}. The latter consist of neutral bubbles of nothing and black strings held together by struts. We have found closed-form solutions with the addition of a magnetic gauge field and constructed the desired geometries \cite{bigpaper}.

\noindent The generalization to rotating solutions will be a crucial step forward and is analogous to twisting the $y$-circle. 

\section{Discussion} 

In this letter, we have discussed a class of smooth bubble solutions in the same regime as non-extremal black strings in Einstein-Maxwell theories. These are good prototypes beyond supersymmetry to test the philosophy of the microstate geometry program for astrophysical black holes.  Moreover the solutions exist for masses and charges where there are no black holes.  They describe solitons in gravity that may correspond to coherent states of quantum gravity.

An important question to address is about stability. It is well-known that gravity with extra dimensions can lead to instabilities. Uncharged black strings have a Gregory-Laflamme instability that forces them to decay to stable black holes \cite{Gregory:1993vy}, while static vacuum bubbles of nothing are semi-classically unstable, but the presence of gauge fields can drastically change this feature. The magnetic black strings in section \ref{sec:BHphase} are free from classical linear instability for $\frac{1}{2}r_\text{S} \leq r_\text{B} \leq r_\text{S}$ \cite{Miyamoto:2006nd}. Extending to $r_\text{B} > r_\text{S}$ shows that the topological stars are classically stable for the full range of parameters \footnote[2]{This follows from a stability analysis by Anindya Dey.}.

It is also interesting to study the physics of the solutions when probed by particles, light rays, or scalar fields \cite{Mayerson:2020tpn}. The difference between the black strings of section \ref{sec:BHphase}, the topological stars, and the usual non-rotating black holes in GR open a new window into black hole physics that cannot be addressed with the current microstate geometries of extremal black holes. Studying the gravitational radiation and Love numbers will be also very interesting for gravitational wave physics. 

In further studies, we would like to investigate physical mechanism for formation of topological stars.  This is an important open question which has to be explored in the context of bubble nucleation in gravity.  

Finally, we point out that the smooth bubbles can behave as heavy particles in nature.  Their sizes are bounded by the radius of the extra dimension and therefore can be microscopic. Their masses are of the order $M \sim \frac{R_y}{\kappa_4^2}$.  Even if we fix the size of the $y$-circle to be at the string scale, these objects have mass of order $10^2 M_P $.  If we allow for multi-bubble objects, we can obtain solitons for any size, from microscopic to macroscopic, with varying choice of mass and charge. It is worth asking whether early universe processes could create stable configurations of massive bubbles that are long-lived as possible new candidates for dark matter.  \vspace{0.25cm}

{\bf{Acknowledgments.}} We are grateful to Iosif Bena, Kim Berghaus, Frederico Bonetti, Anindya Dey, David Kaplan, Daniel R. Mayerson, Nick Warner and Zackary White for interesting conversations and correspondence. The work of IB and PH is supported in part by NSF grant PHY-1820784.

\bibliographystyle{apsrev4-2}
\bibliography{microstates}

\end{document}